\documentclass{Interspeech}
\usepackage{graphicx}
\usepackage{comment}
\interspeechcameraready

\title{Nosey: Open-source hardware for acoustic nasalance}

\author[affiliation={1}]{Maya}{Dewhurst}
\author[affiliation={1}]{Jack}{Collins}
\author[affiliation={1}]{Justin J. H.}{Lo}
\author[affiliation={2}]{Roy}{Alderton}
\author[affiliation={1}]{Sam}{Kirkham}

\affiliation{Linguistics and English Language}{Lancaster University}{United Kingdom}
\affiliation{Department of Language and Communication Science}{City St George's, University of London}{United Kingdom}
\email{\{m.dewhurst1\}\{j.h.lo\}\{s.kirkham\}@lancaster.ac.uk, roy.alderton@city.ac.uk}


\begin{document}

\maketitle

\begin{abstract}
We introduce Nosey (Nasalance Open Source Estimation sYstem), a low-cost, customizable, 3D-printed system for recording acoustic nasalance data that we have made available as open-source hardware (http://github.com/phoneticslab/nosey). We first outline the motivations and design principles behind our hardware nasalance system, and then present a comparison between Nosey and a commercial nasalance device. Nosey shows consistently higher nasalance scores than the commercial device, but the magnitude of contrast between phonological environments is comparable between systems. We also review ways of customizing the hardware to facilitate testing, such as comparison of microphones and different construction materials. We conclude that Nosey is a flexible and cost-effective alternative to commercial nasometry devices and propose some methodological considerations for its use in data collection.
\end{abstract}

\section{Introduction}
\label{sec:introduction}

Nasality is an integral part of human speech that underpins many important questions in phonetics and phonology \cite{hajek2022, ohala1993}. Achieved by velum lowering, nasality can spread from phonologically nasal segments to nearby oral segments via coarticulation in language-, dialect- and speaker-specific ways \cite{bongiovanni2021, clumeck_patterns_1976, cunha2024, pouplier2024}. Such variation in coarticulation has been said to be a key driver of sound change in the development of nasal vowels \cite{beddor_coarticulatory_2009}. Beyond the segmental level, nasality and denasality as voice qualities have been known to attract social evaluation \cite{crosby2023, pittam1987, waaramaa2021}. In clinical settings, atypical displays of nasality play a role in the diagnosis of speech and language disorders \cite{fletcher-etal1999, watterson-etal1993}.

Addressing these issues depends on the ability to reliably measure nasality and compare nasality between speakers. Neither, however, has proven straightforward. Acoustic correlates of nasality in vowels, for example, include A1--P0, F1 bandwidth and spectral tilt \cite{styler_acoustical_2017}. Even the relatively robust A1--P0 can fail, especially for high vowels, when the harmonics associated with both the oral formant and the nasal pole coincide \cite{carignan_practical_2021}. Baseline and range variation in these acoustic features also adds to the challenge of comparing their values across speakers \cite{styler_acoustical_2017}.

An alternative means of investigating (vowel) nasality, long established in the clinical domain and recently gaining popularity in phonetic research, is nasometry. Rather than inferring nasality from spectral characteristics, this method relies on isolating acoustic radiations from the nasal and oral cavities and capturing them using separate microphones. The most common measure used to quantify the output from nasometry is nasalance, which is the ratio between the recorded nasal sound pressure level and the total oral and nasal sound pressure levels \cite{fletcher-etal1974}. A typical calculation of nasalance is

\begin{equation}
\text{Nasalance} = \frac{A_n}{A_n + A_o} \cdot 100
\label{eq:nasalance}
\end{equation}
\\
where $A_n$ and $A_o$ represent acoustic energy captured by the nasal and oral microphones respectively, with the ratio scaled to a percentage value.

Nasalance can reliably separate phonologically oral and nasal vowels, and has been used in various phonetic studies of nasality and nasal coarticulation for this purpose (e.g. \cite{carignan2018, lo2025}). In the clinical realm, normative values of nasalance are typically developed from read materials with varying proportions of nasal sounds, in order to infer and diagnose atypical patterns of nasal airflow (e.g. \cite{brunnegard2009, lehes2018}).

\subsection{Recording nasalance data}

At present, nasometry is typically carried out using commercially available devices. While differing in specific design, existing systems share the common premise of mounting two microphones on either side of an acoustic baffle that is positioned between the nose and the mouth. Systems typically also incorporate microphone pre-amplification and analog-to-digital conversion. Comparisons of nasalance systems have shown a high degree of agreement across different commercial hardware solutions \cite{deboer2014, bressmann-tang2024}. While these offer a user-friendly solution to nasalance data collection, they suffer from a number of drawbacks, such as lack of customizability, a frequent inability to conduct additional analog-domain signal processing, and their relatively high cost, especially for multi-fieldworker contexts where many devices are required.

In the face of these challenges, the `earbuds' method has emerged as a cheap and highly accessible alternative to traditional nasometry. This involves placing one earbud from a pair of earphones below the nostril and the other earbud outside the mouth \cite{stewart-kohlberger2017}. The earbuds can be plugged into a microphone jack, and when incoming sound passes through the speaker coil, it generates a magnetic force that moves the diaphragm (analogous to how a microphone works). The recordings produced tend to be of limited quality, and subsequent evaluation has shown that they are adequate for broad comparisons of average nasality, but are substantially less fine-grained than commercial nasometers at capturing smaller systematic differences \cite{carignan2024}.

We address these issues here by reporting the design and assessment of open-source, 3D-printable hardware for collecting acoustic nasalance data. We note that our system does not aim to fully replace the functionality of commercial nasometers, which may have additional data collection, signal processing and data analysis tools, and also provide customer support. Instead, we believe that our device complements the available commercial offerings, and also provides a testing platform by allowing greater customization of nasometry hardware, including the use of different microphones and nasometer baffle sizes.

\section{Open-source nasalance hardware}

\subsection{Motivations}

We report the design of an open-source, customizable, 3D-printed nasalance system. Our aims were to develop a reliable system that is easy to use, cheap to manufacture and provides additional flexibility as a testing and development platform. For example, our system allows for customized microphone selection and conventional microphone-level analog outputs, which also facilitates custom analog-domain signal processing (e.g. filtering), and analog-to-digital conversion. This opens up the possibility of using higher-quality pre-amplifiers and converters in the signal chain and better integration into multichannel acoustic/articulatory experiments. We provide guidance on modifying the design and all files necessary to 3D print the components in an online GitHub repository.\footnote{\url{https://github.com/phoneticslab/nosey}\\All versioned releases are archived at:\\\url{https://doi.org/10.5281/zenodo.15543852}}

\subsection{3D-printed hardware}

\begin{figure}[t]
  \centering
  \includegraphics[width=\linewidth]{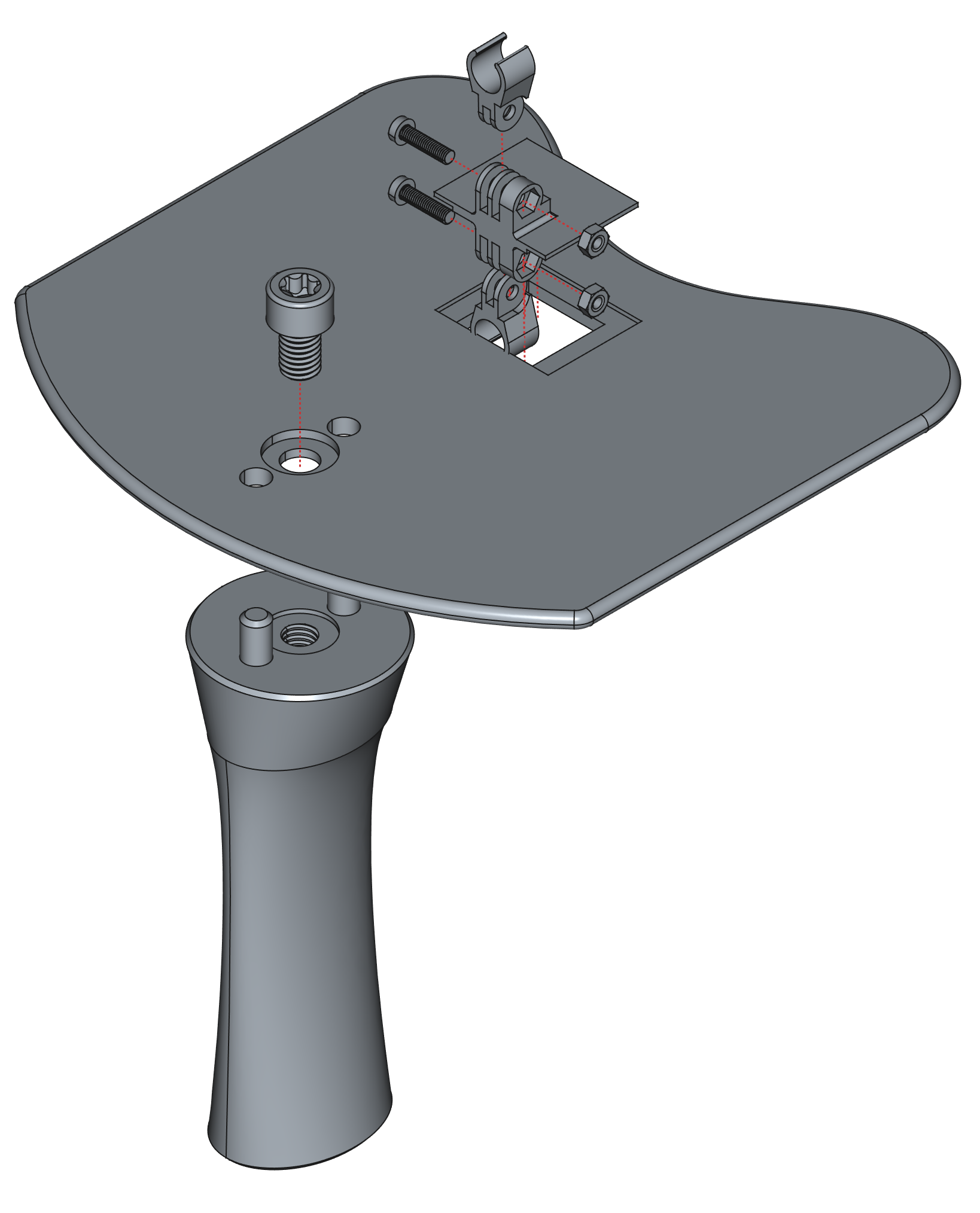}
  \caption{Exploded view of 3D model for the nasalance system.}
  \label{fig:exploded}
\end{figure}

Figure \ref{fig:exploded} shows an exploded view of the 3D model comprising three major components: (i) handle; (ii) baffle; (iii) dual microphone clip, which we review in the rest of this section.

\begin{figure}[t]
  \centering
  \includegraphics[scale=0.18]{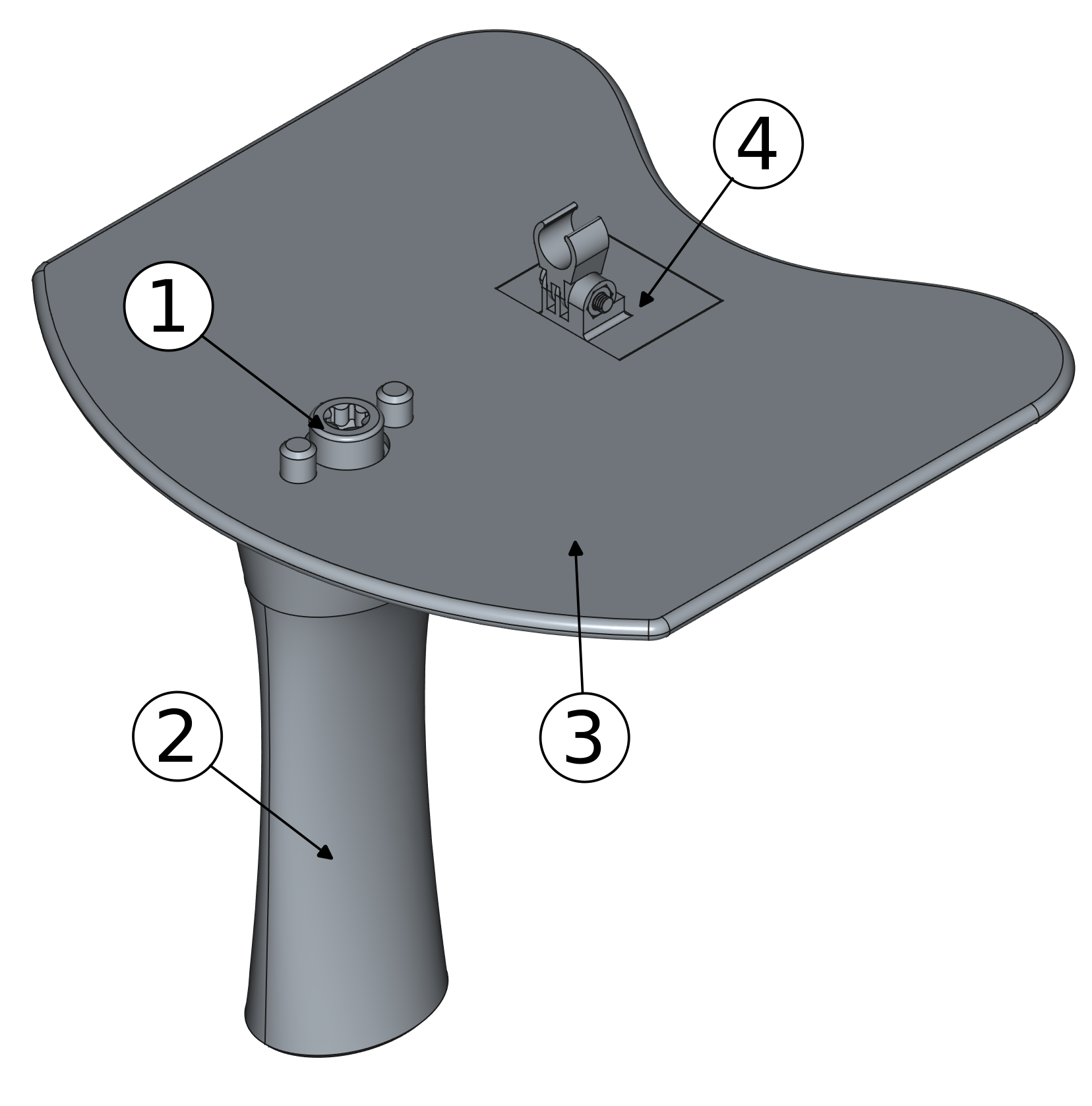}
  \caption{Assembled view of 3D model for the nasalance system.}
  \label{fig:assembly}
\end{figure}

Figure \ref{fig:assembly} shows an assembled view of the 3D model design. The bolt at (1) fixes the handle (2) to the baffle (3). The removable microphone clip at (4) slots into the baffle and can be replaced with alternative versions to allow swapping different microphones and/or clips that are positioned closer to or further away from the edge of the baffle.

\subsection{Microphones and clips}

\begin{figure}[t]
  \centering
  \includegraphics[scale=0.25]{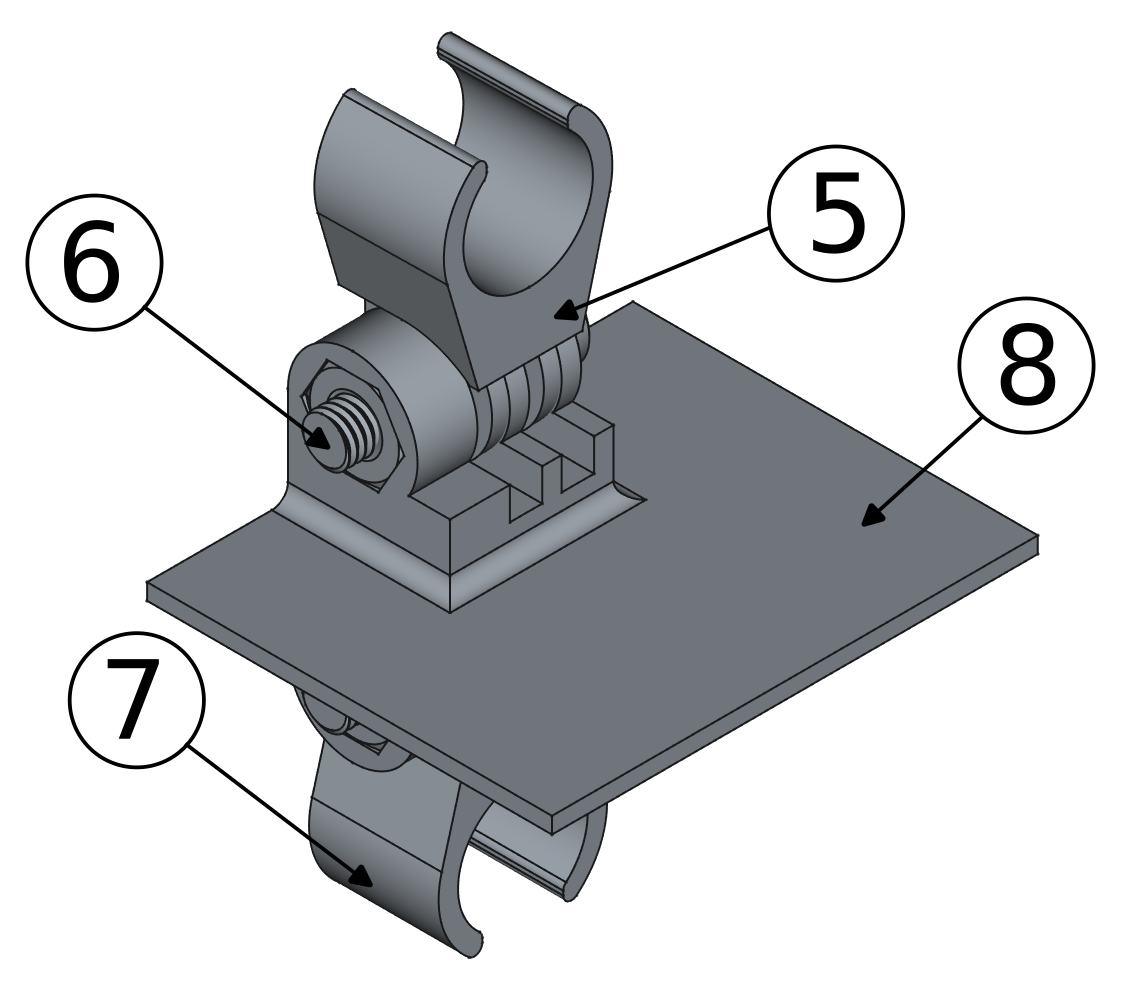}
  \caption{3D model of the removable, articulated, dual microphone holder.}
  \label{fig:mic}
\end{figure}

Figure \ref{fig:mic} shows the 3D diagram of the articulated dual microphone holder. Label (8) shows the base that attaches to the baffle, while (5) and (7) are the microphone clips that attach to the base plate at (6). Note that the microphone clips in (5) and (7) are removable and can be replaced with alternative clips to facilitate the use of different microphones.

The present microphone clips were developed to hold two AKG CK99L microphones, which are lavalier-format, electret condenser microphones, with a cardioid polar pattern and frequency bandwidth of \qty{0.15}{}--\qty{18}{\kilo\hertz}. The front of the clip sits approximately \qty{32}{\milli\metre} away from the front of the baffle; note that this does not include the distance that the microphone extends from the front of the clip. The microphone plate can be adapted for different distances by adapting the FreeCAD models included in the repository.

\subsection{Printing and other components}

A workable model can be achieved using a relatively low-cost 3D printer. For example, we have successfully printed models using the single-extruder MakerBot Replicator+ using PLA plastic. In such cases, the PLA plastic is not food-safe and will absorb moisture, such as saliva and respiratory particles, leading to the growth of bacteria. To lessen the effect of this, the 3D printed baffle and handle can be dipped in food-safe epoxy. An additional measure could include the use of a non-absorbent, disposable barrier.

In addition to the 3D-printed components, the system also requires the following nuts and bolts to assemble the parts:

\begin{enumerate}
    \item M3 nut and bolt (\SI{12}{\milli\metre} long) $[\times2]$
    \item M8 bolt (\SI{12}{\milli\metre} long) $[\times1]$
\end{enumerate}

These bolts are used to fix the baffle to the handle (M8 bolt) and to attach the microphone clips to the removable mounting plate (M3 nut / bolt). This also facilitates the articulated microphone clip, which can be angled by loosening the bolts.

\subsection{Customization}

\begin{figure}[t]
  \centering
  \includegraphics[width=\linewidth]{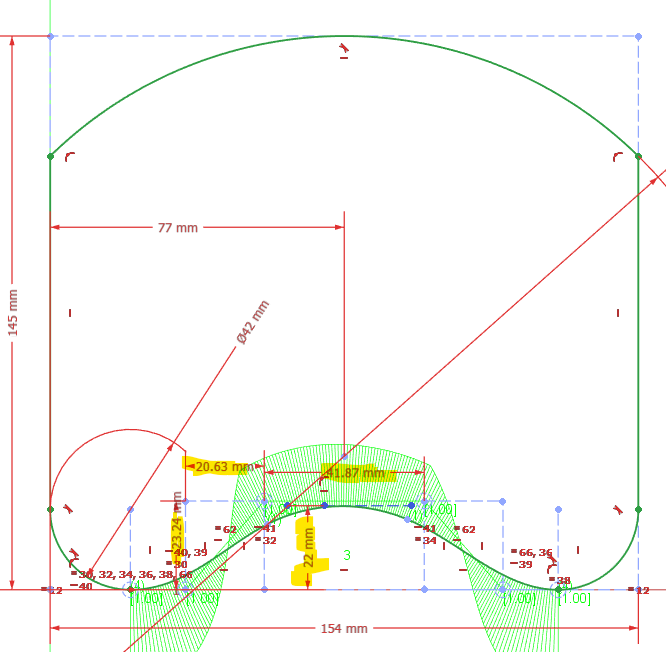}
  \caption{Baffle sketch with four highlighted measurements responsible for changing the shape of the front of the baffle.}
  \label{fig:baffle_edit}
\end{figure}

\begin{figure}[t]
  \centering
  \includegraphics[width=\linewidth]{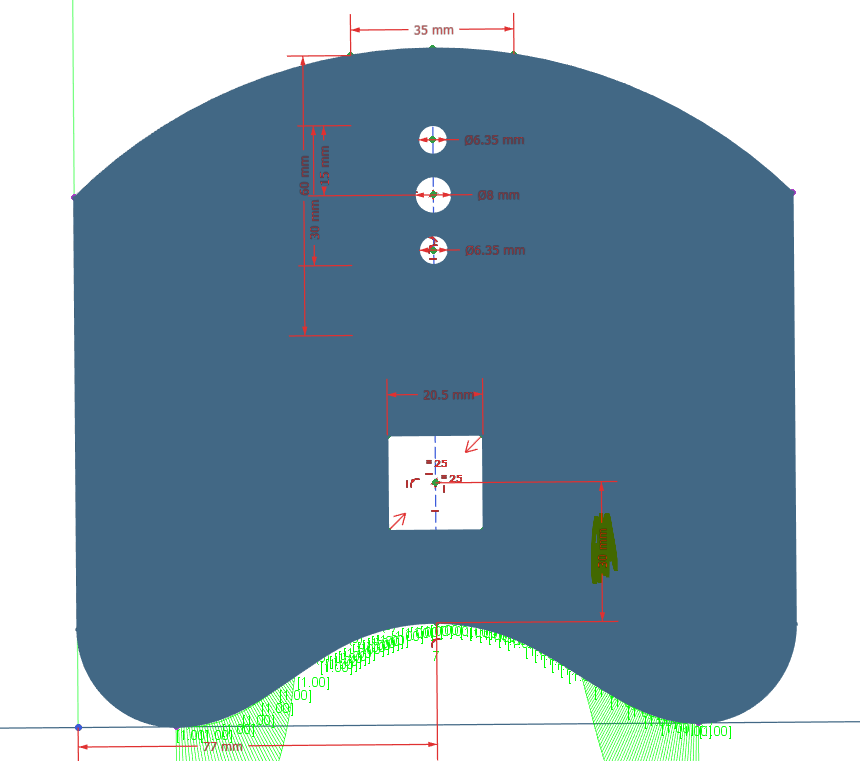}
  \caption{Baffle sketch with highlighted measurement controlling the distance of the cutout to the front of the baffle.}
  \label{fig:cutout_mod}
\end{figure}

We provide editable FreeCAD models and 3D Manufacturing Format files for all model components. The repository includes details on how to modify the baffle shape and distance of the microphone cutout in the baffle. The shape of the baffle can be edited by modifying the highlighted measurements in Figure \ref{fig:baffle_edit}, and the position of the cutout for the microphone can be edited by modifying the measurements in Figure \ref{fig:cutout_mod}. Please see the online repository for further details on customization.

\section{Comparison with commercial system}

\subsection{Overview}

We assessed the performance of Nosey by comparing it to a commercial nasometer: the Nasality Microphone developed by icSpeech (Canterbury, UK). The basic designs of the two systems are very similar (and also very similar to other commercial devices), with two small microphones separated by an acoustic baffle, which is mounted to a handle. Participants hold the device so that the baffle rests against the skin between the nose and the top lip, separating nasal and oral airflow. The purpose of this comparison is to establish the performance of different systems in terms of within-speaker differences in patterns of nasal coarticulation across various phonological environments.

\subsection{Data collection}

We recorded data from two male speakers in their early 30s (authors R.A. and J.L.). R.A. speaks Standard Southern British English (SSBE) as L1, while J.L. speaks SSBE with minor influences from L1 Hong Kong Cantonese. Both speakers read a list of 60 English monosyllabic words containing five vowels (the lexical sets \textsc{kit, dress, trap, strut, face} \cite{wells1982}) in a range of phonological environments that facilitate varying degrees of anticipatory and carry-over nasal coarticulation (e.g. \emph{bin, mid, bend, bent}; see \cite{cunha2024, pouplier2024, lo2025}). For each device, each speaker recorded six repetitions of the word list in randomized order, presented one at a time in a fixed carrier phrase (`I say \_\_\_ for us') via PowerPoint.

Recording was completed in Audacity (v3.7.0) with a constant recording level. The icSpeech Nasality Microphone was connected to the recording laptop via a USB connection, while the two microphones of Nosey were connected via a Focusrite Scarlett 2i2 audio interface and set to equal gains. We removed the wind shields from the icSpeech microphones to eliminate this possible confound from our comparison.

\subsection{Data processing and analysis}

Recordings were automatically transcribed and force-aligned using the Penn Phonetics Lab Forced Aligner (P2FA) \cite{yuan2008}. Intensity extraction for nasalance calculation was also automated, using the \texttt{speakr} package \cite{coretta2025} in R \cite{rcoreteam2025} and a Praat script \cite{boersma2025} written by author M.D. Intensity was extracted at the midpoint of each vowel token, and nasalance was calculated using Equation \ref{eq:nasalance}.

All statistical analyses were conducted in R. To investigate nasalance differences across phonological environments and recording systems, for each speaker we first carried out linear regression with an interaction of phonological environment and system, and a control term for vowel. Variables were deviation-coded for comparison against the grand mean. We used the \texttt{emmeans} package \cite{lenth2025} to calculate the estimated marginal means (EMMs) of nasalance for each combination of system and environment. From these EMMs, we performed pairwise comparisons of how each system captured differences in nasalance between environments. All results were interpreted at $\alpha = .05$, after Bonferroni adjustment was applied to \textit{p}-values to control family-wise error rates. A significant non-zero estimate would suggest that the two systems capture phonological contrasts in nasalance at different magnitudes.

\subsection{Results}

Regression analyses for the two speakers showed significant differences in nasalance between the two systems at the midpoint of the vowel (both $p < .01$). In both speakers, Nosey displays consistently higher nasalance rates than the icSpeech nasometer. For Speaker 1, the average difference between systems is an increase of 22\% when using Nosey ($SE = 0.0030, t = 7.37, p < .01$) and for Speaker 2, Nosey increases nasalance by 21\% ($SE = 0.0033, t = 6.36, p < .01$). This is likely due to the manual aspect of microphone calibration for Nosey, where gain settings must be set by the user. We discuss ways to address this in Section \ref{sec:discuss}.

\begin{figure}[t]
  \centering
  \includegraphics[width=\linewidth]{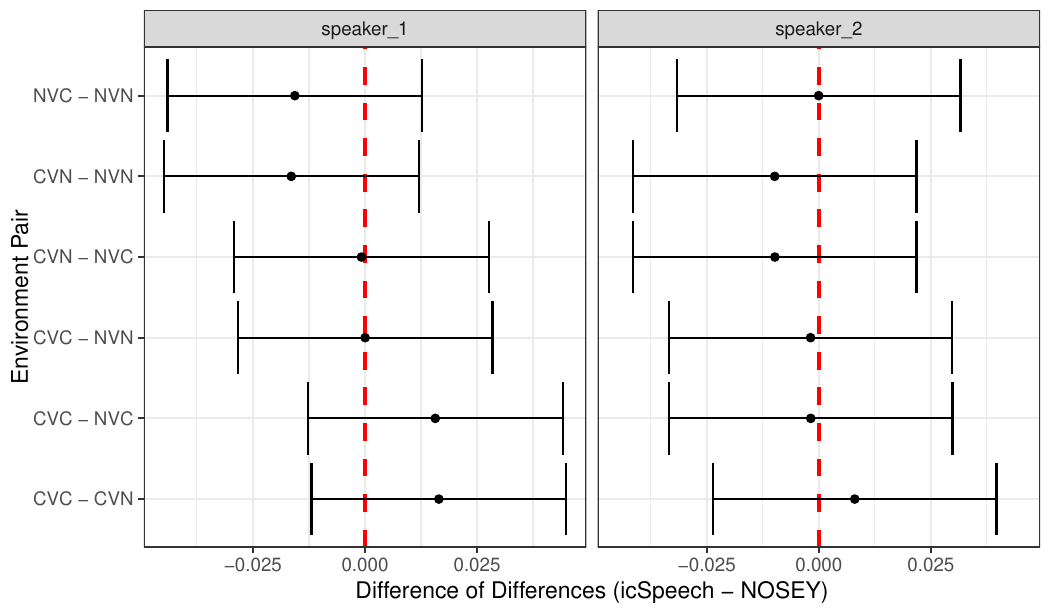}
  \caption{Pairwise comparison of selected phonological environment pairs. Values below zero suggest a greater contrast magnitude for Nosey; values above zero suggest a greater contrast magnitude for icSpeech. An error bar that does not cross zero suggests a significant difference.}
  \label{fig:comparison}
\end{figure}

In terms of differences in nasalance between phonological environments for each system, neither speaker showed significant differences between systems (all adjusted $p > .1$). Figure \ref{fig:comparison} shows means and confidence intervals for the magnitude of the system differences across different phonological environments. There is evidence of higher nasalance for Nosey in some conditions, but the confidence intervals all cross zero. This indicates that neither speaker displayed systematically different magnitudes of nasalance contrast between Nosey and the icSpeech nasometer.

These findings indicate that raw nasalance rates from Nosey are not strictly comparable to the icSpeech nasometer. That said, under the conditions tested here, we can infer that Nosey \textit{can} be relied upon to represent magnitudes of contrasts in nasalance across phonological environments comparable to those reported using commercially available devices. Although further work should explore a larger pool of speakers and take time-varying nasalance rates into account, our results point to Nosey's viability as a flexible, cost-effective alternative to commercial nasometry devices for capturing vowel nasalance across phonological environments.

\section{Discussion}
\label{sec:discuss}

\subsection{Summary and limitations}

We have outlined the development and release of open-source hardware for acoustic nasalance data collection in a way that is accessible and customizable. While a comprehensive validation of Nosey is beyond the scope of this paper, we show that performance is comparable with one commercial system. Raw nasalance is higher with Nosey, but the important comparison of nasalance between phonological environments shows high similarity between systems. 

One possible source of the raw nasalance differences is variation in microphone placement and microphone gain settings. We did not conduct a comprehensive investigation of these factors, but they can easily be addressed due to Nosey's flexible design. For example, microphone placement can be modified by adapting the location of the microphone plate in the FreeCAD model included in the repository. The system also permits full specification of recording settings as it can be used with any audio interface and/or hardware signal processing devices. In future work, we plan to test specific parameter settings that capture the precise patterns shown in commercial systems.

\subsection{Future development}

There are a number of areas for future investigation that are enabled by the Nosey system. First, a productive experiment would be a comparison of different microphone polar patterns. While an omnidirectional pattern provides the flattest frequency response, this makes microphone separation more challenging and significantly increases the likelihood of cross-signal bleed. In our evaluation, we used microphones with a cardioid polar pattern, but it would be useful to compare this with hypercardioid microphones or develop custom microphones with highly directional characteristics. An advantage of our hardware platform is that it facilitates the comparison of different microphones, while keeping the rest of the hardware constant.

Second, it would be helpful to compare the effect of microphone distance on nasalance calculations. In the above sections, we provide guidance on the 3D model measurements that can be changed to modify microphone placement. Third, we have not evaluated the acoustical properties of different baffle materials. For example, it may be desirable to test metal baffle plates, which have a higher mass per unit area and should have better acoustic isolation. This obviously comes at the added complexity of fabricating metal components versus the relative ease of 3D printing plastic, so it is likely that plastic will remain the most popular option. Finally, a valuable topic of investigation would be a comprehensive comparison between Nosey and a wider range of different commercial systems.

\section{Acknowledgements}

This research was supported by UKRI/ESRC Doctoral Fellowship ES/P000665/1 to M.D. and UKRI/AHRC fellowship AH/Y002822/1 to S.K.

\bibliographystyle{IEEEtran}
\bibliography{bibliography.bib, new_refs.bib}

\begin{thebibliography}{10}
\providecommand{\url}[1]{#1}
\csname url@samestyle\endcsname
\providecommand{\newblock}{\relax}
\providecommand{\bibinfo}[2]{#2}
\providecommand{\BIBentrySTDinterwordspacing}{\spaceskip=0pt\relax}
\providecommand{\BIBentryALTinterwordstretchfactor}{4}
\providecommand{\BIBentryALTinterwordspacing}{\spaceskip=\fontdimen2\font plus
\BIBentryALTinterwordstretchfactor\fontdimen3\font minus
  \fontdimen4\font\relax}
\providecommand{\BIBforeignlanguage}[2]{{%
\expandafter\ifx\csname l@#1\endcsname\relax
\typeout{** WARNING: IEEEtran.bst: No hyphenation pattern has been}%
\typeout{** loaded for the language `#1'. Using the pattern for}%
\typeout{** the default language instead.}%
\else
\language=\csname l@#1\endcsname
\fi
#2}}
\providecommand{\BIBdecl}{\relax}
\BIBdecl

\bibitem{hajek2022}
J.~Hajek, ``Nasals and nasalization,'' in \emph{Manual of Romance Phonetics and
  Phonology}, C.~Gabriel, R.~Gess, and T.~Meisenburg, Eds.\hskip 1em plus 0.5em
  minus 0.4em\relax Berlin: De Gruyter, 2022, pp. 215--241.

\bibitem{ohala1993}
J.~J. Ohala, ``Coarticulation and phonology,'' \emph{Language and Speech},
  vol.~36, no. 2, 3, pp. 155--170, 1993.

\bibitem{bongiovanni2021}
S.~Bongiovanni, ``Acoustic investigation of anticipatory vowel nasalization in
  a {Caribbean} and a non-{Caribbean} dialect of {Spanish},'' \emph{Linguistics
  Vanguard}, vol.~7, no.~1, p. 20200008, 2021.

\bibitem{clumeck_patterns_1976}
H.~Clumeck, ``Patterns of soft palate movements in six languages,''
  \emph{Journal of Phonetics}, vol.~4, no.~4, pp. 337--351, 1976.

\bibitem{cunha2024}
C.~Cunha, P.~Hoole, D.~Voit, J.~Frahm, and J.~Harrington, ``The physiological
  basis of the phonologization of vowel nasalization: A real-time {MRI}
  analysis of {American} and {Southern} {British} {English},'' \emph{Journal of
  Phonetics}, vol. 105, p. 101329, 2024.

\bibitem{pouplier2024}
M.~Pouplier, F.~Rodriquez, J.~J.~H. Lo, R.~Alderton, B.~G. Evans, E.~Reinisch,
  and C.~Carignan, ``Language-specific and individual variation in anticipatory
  nasal coarticulation: A comparative study of {American} {English}, {French},
  and {German},'' \emph{Journal of Phonetics}, vol. 107, p. 101365, 2024.

\bibitem{beddor_coarticulatory_2009}
P.~S. Beddor, ``A coarticulatory path to sound change,'' \emph{Language},
  vol.~85, no.~4, pp. 785--821, 2009.

\bibitem{crosby2023}
D.~Crosby, ``\emph{Oppa-Ng Gamsahamnita-Ng$\sim\sim\sim$}: The phonetics of
  nasal cuteness in {Korean} aegyo,'' Ph.D. dissertation, University of South
  Carolina, 2023.

\bibitem{pittam1987}
J.~Pittam, ``Listeners' evaluations of voice quality in {Australian} {English}
  speakers,'' \emph{Language and Speech}, vol.~30, no.~2, pp. 99--113, 1987.

\bibitem{waaramaa2021}
T.~Waaramaa, P.~Lukkarila, K.~J{\"a}rvinen, A.~Geneid, and A.-M. Laukkanen,
  ``Impressions of personality from intentional voice quality in
  {Arabic}-speaking and native {Finnish}-speaking listeners,'' \emph{Journal of
  Voice}, vol.~35, no.~2, pp. 326.e21--326.e28, 2021.

\bibitem{fletcher-etal1999}
S.~G. Fletcher, F.~Mahfuzh, and H.~Hendarmin, ``Nasalance in the speech of
  children with normal hearing and children with hearing loss,'' \emph{American
  Journal of Speech-Language Pathology}, vol.~8, no.~3, pp. 241--248, 1999.

\bibitem{watterson-etal1993}
T.~Watterson, S.~C. McFarlane, and D.~S. Wright, ``The relationship between
  nasalance and nasality in children with cleft palate,'' \emph{Journal of
  Communication Disorders}, vol.~26, no.~1, pp. 13--28, 1993.

\bibitem{styler_acoustical_2017}
W.~Styler, ``On the acoustical features of vowel nasality in {English} and
  {French},'' \emph{The Journal of the Acoustical Society of America}, vol.
  142, no.~4, pp. 2469--2482, 2017.

\bibitem{carignan_practical_2021}
C.~Carignan, ``A practical method of estimating the time-varying degree of
  vowel nasalization from acoustic features,'' \emph{The Journal of the
  Acoustical Society of America}, vol. 149, no.~2, pp. 911--922, 2021.

\bibitem{fletcher-etal1974}
S.~G. Fletcher, I.~Sooudi, and S.~D. Frost, ``Quantitative and graphic analysis
  of prosthetic treatment for ``nasalance" in speech,'' \emph{Journal of
  Prosthetic Dentistry}, vol.~32, no.~3, pp. 284--291, 1974.

\bibitem{carignan2018}
C.~Carignan, ``Using ultrasound and nasalance to separate oral and nasal
  contributions to formant frequencies of nasalized vowels,'' \emph{Journal of
  the Acoustical Society of America}, vol. 143, no.~5, pp. 2588--2601, 2018.

\bibitem{lo2025}
J.~J.~H. Lo, ``Nasal coarticulation in {Lombard} speech,'' \emph{Speech
  Communication}, p. 103205, 2025.

\bibitem{brunnegard2009}
K.~Brunneg{\r{a}}rd and J.~{van Doorn}, ``Normative data on nasalance scores
  for {Swedish} as measured on the {Nasometer}: Influence of dialect, gender,
  and age,'' \emph{Clinical Linguistics \& Phonetics}, vol.~23, no.~1, pp.
  58--69, 2009.

\bibitem{lehes2018}
L.~Lehes, R.~Horn, P.~Lippus, M.~Padrik, P.~Kasen{\~o}mm, and T.~Jagom{\"a}gi,
  ``Normative nasalance scores for {Estonian} children,'' \emph{Clinical
  Linguistics \& Phonetics}, vol.~32, no.~11, pp. 1054--1066, 2018.

\bibitem{deboer2014}
G.~{De Boer} and T.~Bressmann, ``Comparison of nasalance scores obtained with
  the {N}asometers 6200 and 6450,'' \emph{The Cleft Palate Craniofacial
  Journal}, vol.~51, no.~1, pp. 90--97, 2014.

\bibitem{bressmann-tang2024}
T.~Bressmann and B.~H.~Y. Tang, ``Differences in nasalance scores obtained with
  different nasometer headsets,'' \emph{Clinical Linguistics \& Phonetics}, pp.
  1--11, 2024.

\bibitem{stewart-kohlberger2017}
J.~Stewart and M.~Kohlberger, ``Earbuds: A method for analyzing nasality in the
  field,'' \emph{Language Documentation \& Conservation}, vol.~11, no.~1, pp.
  49--80, 20167.

\bibitem{carignan2024}
C.~Carignan, ``Ground-truth validation of the ``earbuds method" for measuring
  acoustic nasalance,'' \emph{Journal of the Acoustical Society of America},
  vol. 156, no.~2, pp. 851--864, 2024.

\bibitem{wells1982}
J.~C. Wells, \emph{Accents of {E}nglish: Volumes 1--3}.\hskip 1em plus 0.5em
  minus 0.4em\relax Cambridge: Cambridge University Press, 1982.

\bibitem{yuan2008}
J.~Yuan and M.~Liberman, ``Speaker identification on the {SCOTUS} corpus,''
  \emph{The Journal of the Acoustical Society of America}, vol. 123, pp.
  3878--3878, May 2008.

\bibitem{coretta2025}
\BIBentryALTinterwordspacing
S.~Coretta, \emph{speakr: A Wrapper for the Phonetic Software 'Praat'}, 2025,
  {R} package version 3.2.4. [Online]. Available:
  \url{https://CRAN.R-project.org/package=speakr}
\BIBentrySTDinterwordspacing

\bibitem{rcoreteam2025}
\BIBentryALTinterwordspacing
{R Core Team}, \emph{R: A Language and Environment for Statistical Computing},
  R Foundation for Statistical Computing, Vienna, Austria, 2025. [Online].
  Available: \url{https://www.R-project.org/}
\BIBentrySTDinterwordspacing

\bibitem{boersma2025}
\BIBentryALTinterwordspacing
P.~Boersma and D.~Weenink, ``Praat: doing phonetics by computer,'' 2025.
  [Online]. Available: \url{http://www.praat.org/}
\BIBentrySTDinterwordspacing

\bibitem{lenth2025}
\BIBentryALTinterwordspacing
R.~V. Lenth, \emph{emmeans: Estimated Marginal Means, aka Least-Squares Means},
  2025, {R} package version 1.10.7. [Online]. Available:
  \url{https://CRAN.R-project.org/package=emmeans}
\BIBentrySTDinterwordspacing

\end{thebibliography}

\end{document}